\documentclass[
preprint,
 amsmath,amssymb,
aps,
pra,
]{revtex4-2}

\usepackage{graphicx}
\usepackage{dcolumn}
\usepackage{bm}
\usepackage{hyperref}
\usepackage{xcolor}
\usepackage{multirow}
\usepackage{booktabs}
\usepackage{amsmath}

\begin{document}
\preprint{APS/123-QED}

\title{Edge-Stabilized Rotating Flames in a Circular Hele-Shaw Cell}
\author{Xiangyu Nie, Shengkai Wang}
\thanks{Corresponding author: sk.wang@pku.edu.cn (Shengkai Wang)}
\affiliation{SKLTCS, CAPT, School of Mechanics and Engineering Science, Peking University, 5 Yiheyuan Road, Haidian District, 100871, China}
\date{\today}

\begin{abstract}
In this study, we report direct experimental observations of self-sustaining  CH$_4$-air rotating flames formed spontaneously in an unheated, open, circular Hele-Shaw cell. These flames are observed under fuel-rich conditions and exhibit stable traveling-wave patterns, with edge velocities that can significantly exceed the nominal flame speed of the unburned mixture. PLIF measurements across the central plane reveal that the flame front consists of a bibrachial structure, with a diffusion branch gliding along the side edges of the cell and a premixed branch extending into the interior. Complementary numerical simulations suggest that the formation of rotating flames is driven by a dynamic balance between local flame speed and unburned-gas velocity near the cell edges, where both wall heat loss and flow expansion play critical roles in stabilizing the rotation pattern. A parametric study is conducted for various equivalence ratios, flow rates, and gap distances, from which the regime diagrams of flame modes and rotation frequencies are obtained. \color{black}At low flow rates, the rotating state is characterized by a single rotating flame wave, whose rotation frequency increases with flow rate. \color{black}For this type of flames, a semi-empirical model is established to predict their rotation frequencies and shapes as functions of mass flow rate and surface temperature. \color{black}At elevated flow rates, multiple rotating waves appear with approximately equal azimuthal spacing, and the product of the wave number and rotation frequency increases with flow rate. \color{black}Mode transition from rotating flames to steady ring-shaped flames anchored at the burner edges occurs at sufficiently high flow rates, while at sufficiently low flow rates, flame extinction occurs due to thermal quenching. These findings can provide useful guidance for the advancement of micro-combustion technologies.
\end{abstract}

\keywords{Laminar Flame Propagation, Rotating Flame, Flame Stabilization, Hele-Shaw Cell, Heat Loss Effect}
\maketitle
\clearpage

\section{Introduction}
The study of flame dynamics in micro-channels has attracted increasing research attention due to its relevance to the development of micro-combustion technologies \cite{E2022122509, FERNANDEZPELLO2002883, maruta2011micro} and safety protection against unwanted fires and explosions in confined spaces \cite{martinez2019role}. Compared with freely propagating flames, a key difference for flames propagating in confined spaces, such as channels, cells, and micro-combustors, is the presence of thermal and chemical quenching at walls, which significantly modifies the flame dynamics. To better understand these phenomena, experimental investigations of premixed flames propagating in narrow chambers are often performed, where they can be amplified by geometric confinement of the flow \cite{martinez2019role}. 

There have been a myriad of previous studies on the propagation of flames in static mixtures within closed or semi-closed channels, as these flames served as a platform to investigate various forms of instabilities (for example, hydrodynamic \cite{al2019darrieus, shen2019flame, sarraf2018quantitative}, thermal-diffusive \cite{daou2021effect}, and thermal-acoustic \cite{veiga2019experimental} instabilities) under two-dimensional, simplified geometries. However, flame studies in flowing channels have been relatively scarce, in part due to the complications involved in designing suitable micro-channel devices capable of performing such experiments.

Previously, researchers at Tohoku University \cite{kumar2007formation, kumar2007pattern, fan2009experimental, fan2013flame} have used a bottom-heated circular Hele-Shaw cell with well-controlled temperature profiles to investigate various flame patterns and their transition dynamics, opening new pathways for experimental studies on this aspect. Under strong external heating, the wall quenching effects were inhibited, and a negative temperature gradient along the radial direction was established (cold in the center and hot near the cell edges), enabling sustained propagation of flames inside the narrow gap of the Hele-Shaw cell. In particular, an interesting pattern of rotating flames (Pelton-like \cite{fan2009experimental} or spiral-like \cite{kumar2007pattern}) was observed at average wall temperatures around 800 K.

These self-sustained traveling reaction waves are fascinating as they can increase the effective burning rate and total thermal power beyond that of stationary laminar flames, in a manner similar to rotating detonation waves observed in pressure-gain combustion systems \cite{zhou2016progress}. However, to date, the exact range of conditions under which such rotating flames exist remains largely uncertain. For example, to the best of the authors’ knowledge, similar types of rotating flames have not been reported for unheated/cold Hele-Shaw cells.

A similar gap exists in both theoretical and numerical investigations of such rotating flames. Several numerical studies have been conducted for flames inside externally heated circular Hele-Shaw cells, focusing on flame propagation and stability characteristics. For example, Minaev and co-workers \cite{minaev2009splitting, minaev2013oscillating} demonstrated the possibility of oscillatory and rotating flame structures based on thermal-diffusive analysis with prescribed wall-temperature distributions and Poiseuille-type velocity fields, relating their emergence to wall-induced preheating, velocity non-uniformity, and the instability branches of S-shaped stationary solutions. In another study, Fan et al. \cite{fan2009experimental} showed that the rotating pelton-like structures in heated radial microchannels could be quantitatively described by simplified reacting flow models with one-step global chemistry. More recently, Chang and Kang \cite{chang2024numerical} performed three-dimensional simulations for the temporal variations of flame structures in a radial microchannel at high wall temperatures, and numerically observed the formation of stationary, splitting/merging, pelton-like, and sweeping flame patterns under different flow and boundary conditions. However, numerical or theoretical studies of rotating flames in unheated Hele-Shaw cells are scarce.

In the present study, we report direct experimental observation of a type of rotating flames of rich methane-air mixtures that spontaneously formed in an unheated, open, circular Hele-Shaw cell. Unlike flames propagating in closed Hele-Shaw channels or heated circular cells, these rotating flames exhibit an edge-stabilized pattern of traveling waves, with leading flame fronts propagating along the rim of the cell and long tails extending into the gap. The mechanism of rotating flame formation and stabilization is further investigated using complementary numerical simulations.

\section{Methods}
\subsection{Experimental Setup}
The current flame experiments were performed in a 200-mm diameter circular Hele-Shaw cell formed by two parallel plates, as shown in Figure 1. The top plate was made of 2.5 mm thick JGS1 fused quartz, which allowed optical access in the wavelength range of 185-2500 nm. The bottom plate was made of SAE 304 stainless steel with a thickness of 25 mm. The upper quartz plate was suspended by three thin wires and precisely aligned with the bottom plate using a bubble level to an accuracy of 0.1$^{\circ}$, thus ensuring a uniform gap distance between the plates.

The test gas mixture of high-purity fuel (99.99\%-grade CH$_4$) and synthetic air (prepared from 99.999\%-grade O$_2$ and 99.999\%-grade N$_2$) was supplied to the Hele-Shaw cell through a 4-mm-diameter entrance port at the center of the bottom plate. The fuel and air were thoroughly premixed in an in-line static mixer upstream of the entrance port, and their flow rates were accurately controlled by two Alicat MC-series mass flow controllers, with typical uncertainties of 0.1\% and 0.2\%, respectively.

\begin{figure}[ht]
\includegraphics[width=0.8\linewidth]{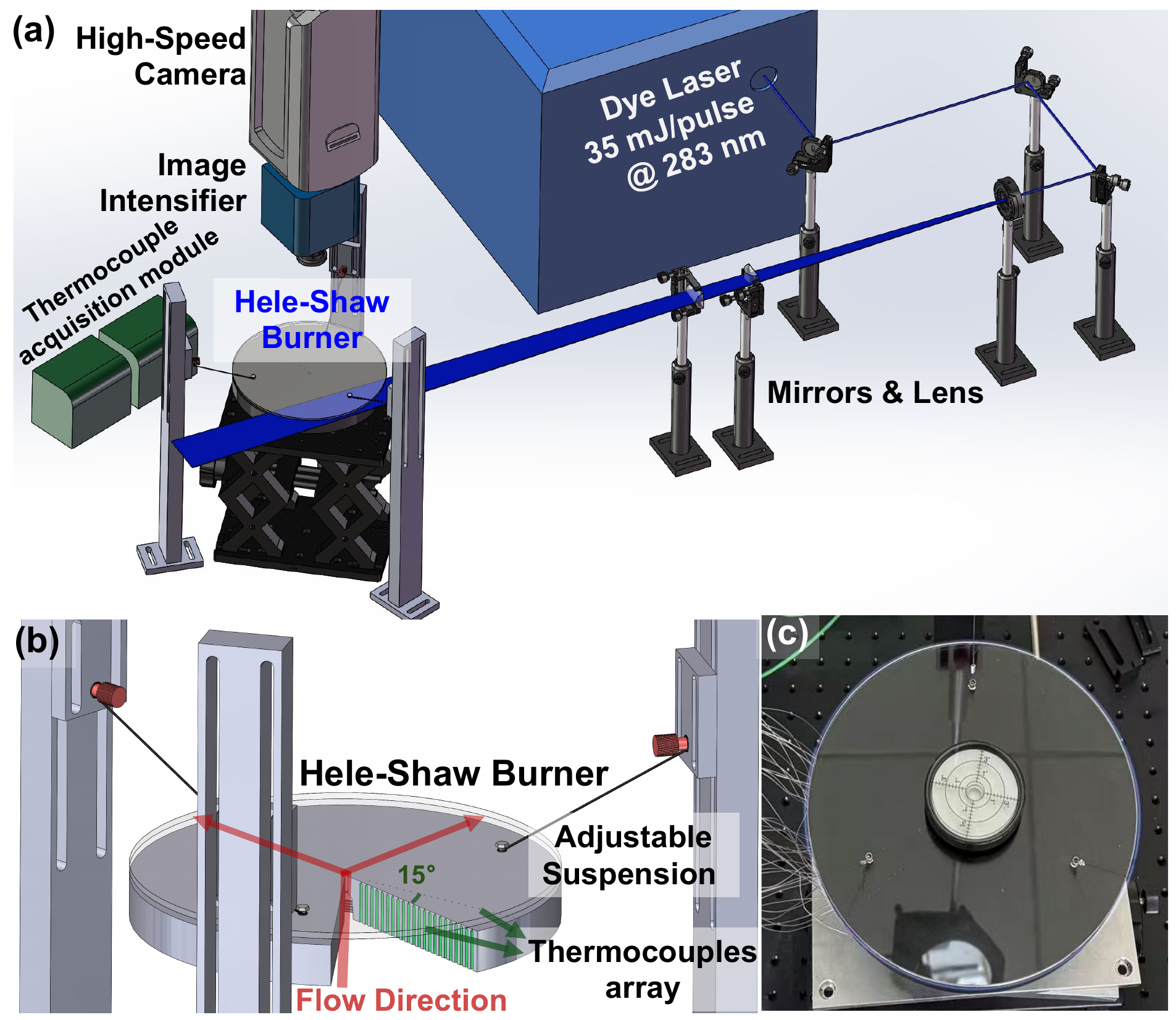}
\caption{(a) Schematic of the current experimental setup. (b) Detailed view of the Hele-Shaw burner. (c) Photo of the burner and the gas supply system.}
\label{fig_1} 
\end{figure}

To accurately measure wall temperatures, two arrays of shielded K-type thermocouples of 0.5 mm tip diameter were mounted on the bottom plate of the Hele-Shaw cell. These thermocouples were flush-mounted on the upper surface of the plate to minimize the disturbance to flames. A total of 44 thermocouples were arranged along two radial lines separated by 15 degrees, providing spatially resolved temperature measurements from 18 to 98 mm along the radial direction at 2 mm intervals. The signals of all thermocouples were continuously acquired, digitally recorded at a sampling rate of 1 Hz, and streamed to a computer in real time via RS-485 communication protocol using three 16-channel data acquisition modules (Rise PK9019). \color{black}The overall 1$\sigma$ uncertainty in the measured temperatures was estimated to be less than 2\%.\color{black}

The flame dynamics and structure were investigated using OH-PLIF and OH* chemiluminescence diagnostics, based on an optical setup as shown in Figure 1(a). For the OH-PLIF measurements, the A-X (1,0) P(1.5)+Q(1.5)+R(2.5) transition cluster of OH radicals at a vacuum wavelength of 282.997 nm was excited by a nanosecond-pulsed tunable dye laser (LIOP-TEC, LiopStar-N) operating at 10 Hz. The dye laser was pumped by a Nd:YAG laser (InnoLas, SpitLight 2000-10) at 532 nm to generate coherent radiation at 566 nm, which was subsequently frequency-doubled to generate 30-mJ pulses at 283 nm. The excitation laser beam was reshaped into a thin planar sheet of less than 0.3 mm thickness using three cylindrical fused-silica lenses with focal lengths of -25 mm, 500 mm, and 500 mm, respectively. The laser sheet was aligned to the central plane of the Hele–Shaw burner, illuminating a region of approximately 50 mm width along the edge of the Hele-Shaw burner. The OH-PLIF signal was spectrally filtered over 300 - 320 nm and recorded with an intensified high-speed CMOS camera (Phantom v611 with EyeiTS intensifier) that was gated (with an exposure time of 10 $\mu$s) and synchronized with the laser pulses. 5-kHz OH* chemiluminescence measurements were conducted using the same optical setup but with the excitation laser turned off. The pixel resolution of the recorded images was 656 $\times$ 864, with each pixel corresponding to a physical size of 280 $\mu$m $\times$ 280 $\mu$m. The effective spatial resolution, usually limited by the image intensifier, was determined to be approximately 0.3 mm (on the order of 3 pixels) using a geometry calibration target. Further details on the calibration of the current imaging system can be found in the authors' previous work \cite{wang2025self}. 

\subsection{Numerical Simulations}
Complementary numerical simulations were conducted for representative operating conditions using the EBIdnsFOAM solver \cite{bockhorn2012implementation, zirwes2023assessment} within the OpenFOAM platform. The methane-air reaction kinetics was modeled using the DRM-19 reduced mechanism, which comprises 21 species and 84 elementary reactions\cite{kazakov1994reduced}. In addition, laminar flame-speed calculations were performed with Cantera\cite{cantera} based on a one-dimensional reacting-flow configuration. 

\color{black}Specifically, the present numerical simulations are performed with variable-density, multi-species reacting-flow calculations, with governing equations for mass, momentum, species mass fractions, and total sensible enthalpy as follows.
\begin{equation}
\frac{\partial \rho}{\partial t}
+\nabla\cdot(\rho \mathbf{u})=0.
\label{eq:ebi_mass}
\end{equation}

\begin{equation}
\frac{\partial (\rho\mathbf{u})}{\partial t}
+\nabla\cdot(\rho\mathbf{u}\mathbf{u})
=-\nabla p+\nabla\cdot\boldsymbol{\tau}.
\label{eq:ebi_momentum}
\end{equation}

\begin{equation}
\frac{\partial (\rho Y_k)}{\partial t}
+\nabla\cdot\left[\rho(\mathbf{u}+\mathbf{u}_c)Y_k\right]
=\dot{\omega}_k-\nabla\cdot\mathbf{j}_k,
\qquad k=1,\ldots,N_s-1.
\label{eq:ebi_species}
\end{equation}

\begin{equation}
\frac{\partial}{\partial t}
\left[\rho\left(h_s+\frac{1}{2}\mathbf{u}\cdot\mathbf{u}\right)\right]
+\nabla\cdot
\left[\rho\mathbf{u}
\left(h_s+\frac{1}{2}\mathbf{u}\cdot\mathbf{u}\right)\right]
=
-\nabla\cdot\dot{\mathbf{q}}
+\frac{\partial p}{\partial t}
-\sum_{k=1}^{N_s} h_k^\circ \dot{\omega}_k,
\label{eq:ebi_energy}
\end{equation}

In these equations, $\rho$ is the gas density, $\mathbf{u}$ is the flow velocity, 
$p$ is the gas pressure, and $\boldsymbol{\tau}$ is the viscous stress tensor. $Y_k$, $\dot{\omega}_k$, $\mathbf{j}_k$, and $h_k^\circ$ are the mass fraction, chemical production rate, diffusive mass flux, and formation enthalpy of species k, respectively, and $N_s$ is the total number of species. The correction velocity $\mathbf{u}_c$ is introduced to ensure that the sum of all diffusive mass fluxes is zero. $h_s$ is the sensible enthalpy of the gas mixture, and $\dot{\mathbf{q}}$ denotes the total heat flux vector including thermal conduction and enthalpy transport by species diffusion. The overall heat release rate from chemical reactions is represented by the term $-\sum_{k=1}^{N_s} h_k^\circ \dot{\omega}_k$. An ideal-gas equation of state is assumed, i.e., $\rho={p\overline{M}}/{R_uT}$, where $\overline{M}$ is the mean molar mass of the mixture and $R_u$ is the universal gas constant.\color{black}

The computation domain and boundary conditions of the current simulations are illustrated in Figure 2. \color{black} The primary aim of these simulations is to characterize the local structure at the tip of a rotating flame along the burner edge where the premixed and diffusion branches merge, enabling insight into the stabilization mechanism of the flame in the radial direction without resorting to a full 3D analysis. \color{black} To this end, a simplified computation domain -- a 1-degree quasi-axisymmetric wedge sector -- was adopted. The cross-section of the current computation domain ranges from $r=80$ to $160~\mathrm{mm}$ in the radial direction (the cell radius is $R$ = 100 mm) and from $z=-30$ to $1.5~\mathrm{mm}$ in the vertical direction. Both the internal gap of the Hele-Shaw burner ($0\leq z\leq 1.5~\mathrm{mm}$) and an external region below the bottom plane ($z<0$ mm) were included to capture flow expansion and entrainment near the burner rim. A symmetric setup was adopted to reduce the computation cost by half, with the upper boundary of the domain ($z=1.5~\mathrm{mm}$) defined as the plane of symmetry. A parabolic velocity profile representing fully developed Poiseuille flow was assigned as the inlet boundary condition, and the mass fractions of CH$_4$ and air were set according to the experimental values. The burner surface was modeled as a no-slip wall \color{black}with experimentally determined temperature profiles. \color{black}The rest of the boundary was assigned zero-gradient conditions for velocity, temperature, and species mass fractions. Three representative cases are shown in Fig.~8.

The local flame structure and flow field near the burner edge are examined using the quasi-axisymmetric simulations described above. These calculations represent a radial--vertical cross-section through an established flame wave and are not intended to reproduce the azimuthal motion of the rotating pattern. 

\begin{figure}[ht]
\includegraphics[width=0.75\linewidth]{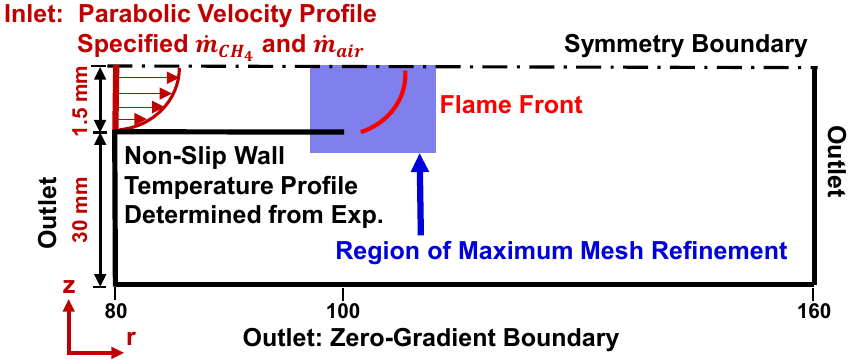}
\caption{Schematic of the computational setup and boundary conditions.}
\label{fig_2} 
\end{figure}

Spatial discretizations were performed using a structured multi-block mesh that was progressively refined toward the burner wall and the flame stabilization region. In the near-wall region ($0\leq z\leq 1.5~\mathrm{mm}$), a mild geometric stretching was applied, resulting in cell sizes of $\Delta z \approx 0.03 - 0.05~\mathrm{mm}$. Along the radial direction, the mesh was locally refined near the burner rim ($x\approx 95$-$110~\mathrm{mm}$), and a minimum radial spacing of $\Delta x \approx 0.05~\mathrm{mm}$ occurred near $x\approx 100~\mathrm{mm}$. Away from this region, the mesh was gradually coarsened toward the far-field boundary, with a maximum cell size of approximately $1.0~\mathrm{mm}$. The expansion ratios between adjacent cells were limited to a few percent to maintain numerical robustness. The entire mesh contained approximately $2.9\times 10^{4}$ cells.

\color{black}
A full 3-D simulation was also conducted to resolve the azimuthal structure of a representative rotating flame. The computation domain was constructed by extending the vertical cross-section in the azimuthal direction by 360 degrees while maintaining the same boundary conditions. Adaptive mesh refinement (AMR) was used to concentrate spatial resolution in the flame region while avoiding excessive computational cost \cite{RETTENMAIER2019100317}. During the calculation, the mesh was re-evaluated every 10 steps based on the local OH mass fraction. Cells satisfying $Y_{\rm OH}>2\times10^{-3}$ were refined, while cells outside the reaction layer were coarsened. Starting from a background mesh size of approximately 0.5 mm, three refinement levels were applied, resulting in a finest grid spacing of about 0.05 mm near the flame front.
\color{black}

\section{Results and Discussion}
\subsection{Representative Flame Structure}

\begin{figure}[ht]
\includegraphics[width=0.9\linewidth]{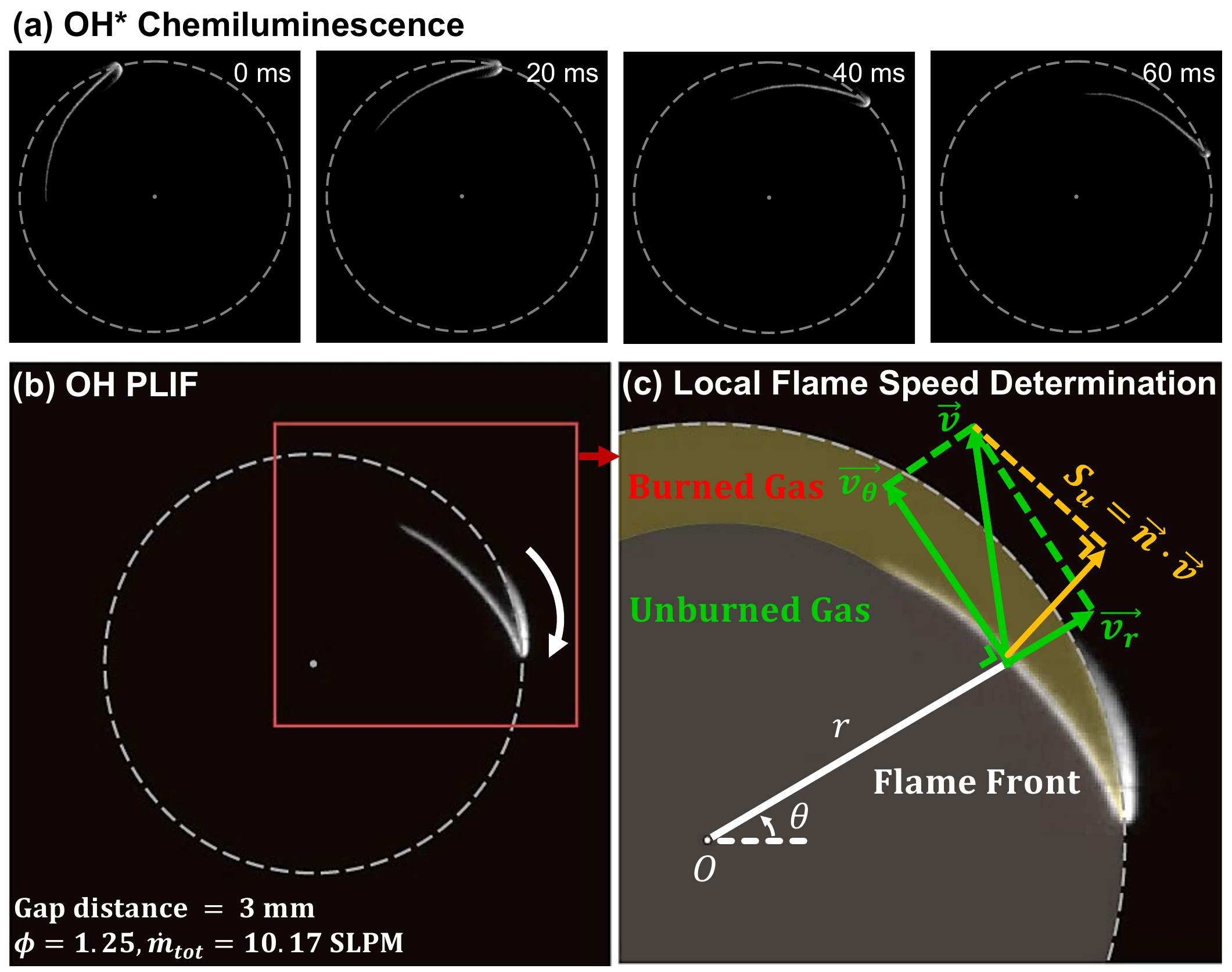}
\caption{Representative images of an edge-stabilized single-wave rotating flame of CH$_4$-air mixture at $\phi$ = 1.25 and $\dot{m}_{tot}$ = 10.17 SLPM. (a) Sequential images of OH chemiluminescence, (b) a representative OH-PLIF image at the same condition and (c) graphic illustration of the local flame speed reconstruction.}
\label{fig_3} 
\end{figure}

Figure 3(a) presents a representative sequence of OH* chemiluminescence images for an edge-stabilized rotating CH$_4$-air flame observed in the unheated circular Hele-Shaw cell at an equivalence ratio $\phi$ = 1.25 and a total mass flow rate $\dot{m}_{tot}$ = 10.17 SLPM. \color{black}The flame followed a stable single-wave rotating pattern, in which the leading flame front propagated azimuthally along the rim of the cell at a frequency of approximately 3 Hz, while the trailing flame extended obliquely into the interior of the cell. \color{black}In these images, the leading reaction zone appeared to have a T-shaped structure similar to that of triple-flames in stratified mixtures; however, we note that this was merely an artifact caused by spatial overlapping of chemiluminescence signals at different heights outside the cell. This artifact was eliminated in the OH-PLIF measurements that selectively illuminated the central horizontal plane, as illustrated in Figure 3(b), and a bibrachial flame structure with well-defined reaction fronts was clearly observed near the edge of the cell.

\subsection{Radial Distribution of Flame Speed under Heat Loss}

For an edge-stabilized rotating flame, the local flame speed $S_u$ is calculated from the observed flame front location using a polar coordinate system $(r,\theta)$ in the co-rotating frame of reference, as shown in Figure 3(c). The flame front location is represented as $r =r(\theta)$, where $r$ is the radial distance from the burner center and $\theta$ is the azimuthal angle. The unit normal vector $\mathbf n$ along the flame front can be expressed as 
\begin{equation}
\mathbf n = \frac{r}{\sqrt{r^2+(r')^2}} \mathbf{e_r} - \frac{r'}{\sqrt{r^2+(r')^2}} \mathbf{e_\theta}
\end{equation}
where $r'=dr/d\theta$; $\mathbf{e_r}$ and $\mathbf{e_\theta}$ represent the radial and transverse unit vectors, respectively. 

\color{black}
The radial component of the flow velocity, $v_r$, can be readily determined from mass conservation under the Poiseuille-flow assumption – a simplified treatment that approximates the unburned gas as axisymmetric radial flow with a parabolic profile in the vertical direction. This yields $v_r(r) = 1.5 \bar{v}_r(r)$ and  $2\pi rH\rho_u(r)\bar{v}_r(r)=\rho_0\dot{V}_{\rm in}$, where $\bar{v}_r $is the gap-averaged radial velocity, $H$ is the gap distance, $\rho_u(r)$ is the local density of the unburned mixture, $\rho_0$ is the density at the reference temperature $T_0=300$ K, and $\dot{V}_{\rm in}$ is the volumetric flow rate at standard temperature and pressure. The density variation of the gas mixture is also considered to account for changes in volumetric flow rate caused by thermal expansion, i.e.,${\rho_u(r)}/{\rho_0} \approx {T_0}/{T_s(r)}$, where $T_s(r)$ is the surface temperature obtained by interpolation of thermocouple measurements. Therefore, the radial velocity along the center plane is

\begin{equation}
v_r(r)= \frac{\dot{3V}_{\rm in}}{4\pi rH} \frac{T_s(r)}{T_0}.
\label{eq:vr_poisseuille}
\end{equation}\color{black}

In the inertial frame of reference, the transverse (azimuthal) velocity of the rotating flame front is:
\begin{equation}
v_{\theta}=2\pi r f,
\end{equation}
where $f$ is the experimentally observed rotation frequency of the flame pattern. \color{black} This apparent velocity results from the coordinate transformation between the laboratory frame and the front-fixed rotating frame. In the laboratory frame, the unburned mixture ahead of the flame front has only a velocity component in the radial direction. \color{black}In the co-rotating frame of reference, the flame front appears stationary, and the velocity vector of the unburned gas can be expressed as $\mathbf{v}= v_r \mathbf{e_r} + v_{\theta}\mathbf{e_\theta}$.

The local flame speed is determined by projecting $\mathbf{v}$ onto the normal direction of the flame front, i.e., 
\begin{equation}
S_\mathrm{u}(r) = \mathbf{v}\cdot\mathbf{n} = \frac{v_r\,r - v_\theta \,r'}{\sqrt{r^2+(r')^2}}.
\label{eq:su}
\end{equation}
For the representative example shown in Fig. 3, the calculated radial distribution of local flame speed is presented in Fig. 4. \color{black}The uncertainty in $S_u$ originates from three sources: (a) the wall temperature $T_s$, (b) the rotation frequency $f$, and (c) the flame-front coordinates extracted from flame images. The overall $1\sigma$-uncertainty in the measured temperatures is estimated to be less than 2\%, which affects the reconstructed flame speed by less than 1\% (primarily through the expansion factor $T_s/T_0$ in the calculation of $v_r$). The uncertainty in the rotation frequency, arising from time-resolved flame-wave tracking of the high-speed (5 kHz) images, is also below 1\%. The dominant uncertainty source is the effective spatial resolution of the optical system (approximately 0.3 mm). Its contribution to the local flame speed uncertainty is estimated via a bootstrapping method: the flame front coordinates are randomly perturbed along the local normal direction with a standard deviation of 0.3 mm, and the local flame speed $S_u$ is evaluated repetitively. From the standard deviation of this distribution, the uncertainty in $S_u$ is estimated to be less than 10\%, as shown in Fig. 4(b), with the maximum uncertainty occurring at the edge of the burner.\color{black}

\begin{figure}[ht]
\includegraphics[width=0.5\linewidth]{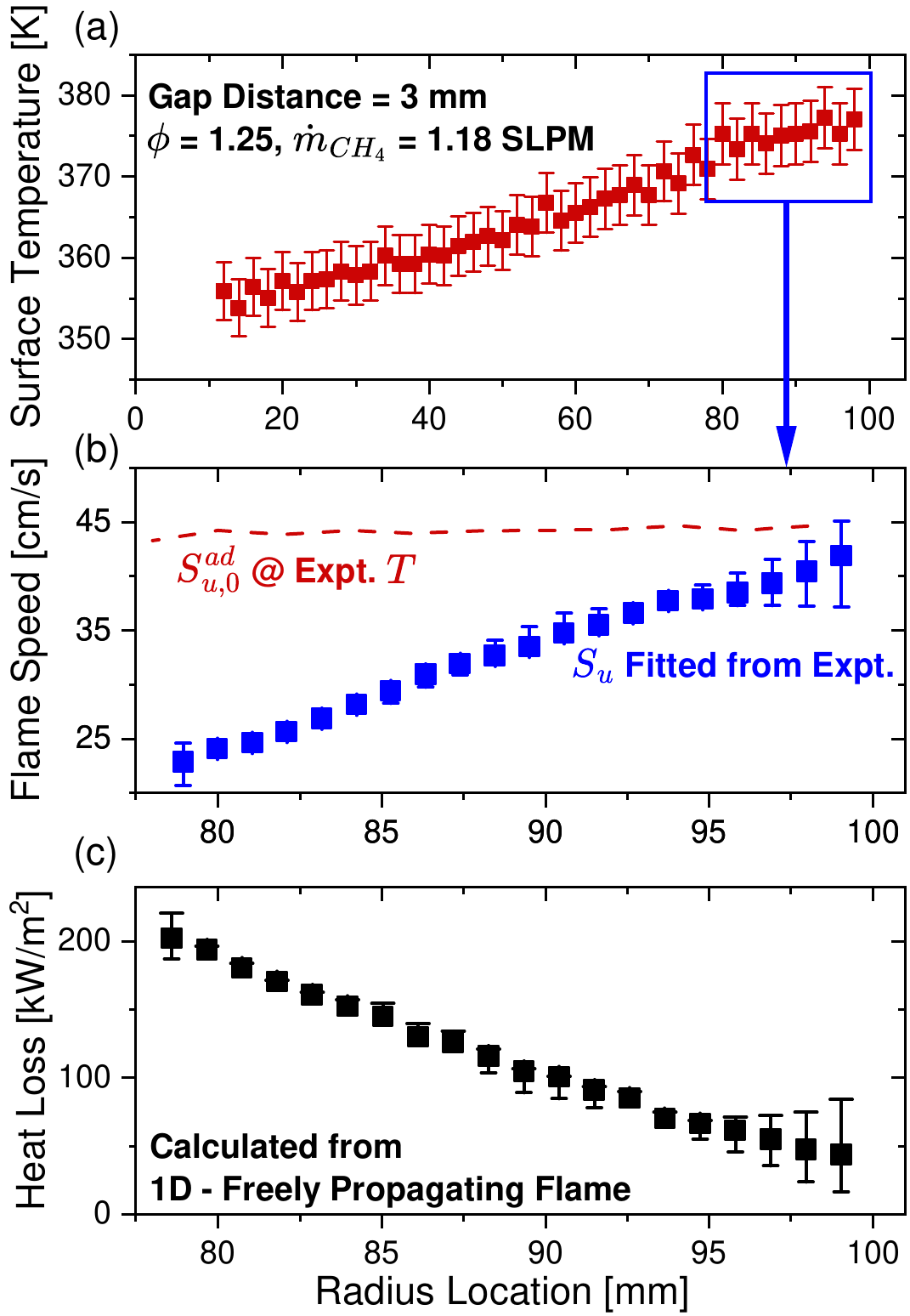}
\caption{Radial distributions of (a) the measured surface temperature of the Hele-Shaw burner, (b) the local flame speed determined from experiment, plotted in comparison with the adiabatic flame speed evaluated at the measured surface temperature, and (c) the effective heat loss rate inferred from the one-dimensional calculation. Results are shown for the representative case of rotating flame in Fig. 3.}
\label{fig_4} 
\end{figure}

It is evident that the flame speed varies significantly with radial location. Near the burner edges, the flame speed approaches the adiabatic flame speed at the corresponding surface temperature, while it decreases rapidly away from the edges. At a radial distance of approximately 20 mm from the edges, the local flame speed drops to about 40\% of the adiabatic value; beyond this point, the flame extinguishes. 

Also shown in Fig. 4 is the radial distribution of the measured surface temperature. Similar to the trend in local flame speed, the surface temperature reaches a maximum at the edges and decreases toward the interior, yielding a positive temperature gradient along the radial direction. However, the relative change in surface temperature is much smaller than that of the local flame speed, and the variation in the observed local flame speed is an order of magnitude greater than the corresponding adiabatic value, suggesting that it is dominated by heat loss through the walls.

To analyze the heat-loss effect, a simplified model is used for the local flame speed, incorporating an effective volumetric heat-loss term $\dot{q}_\mathrm{loss}$. Specifically, the flame speed ${S}_\mathrm{u}^\mathrm{mod}(\dot{q}_\mathrm{loss})$ is evaluated based on a one-dimensional non-adiabatic simulation using EBIdnsFoam, where $\dot{q}_\mathrm{loss}$ is included in the sensible-enthalpy equation. The local value of $\dot{q}_\mathrm{loss}(r)$ is inferred from the experimental value of local flame speed, i.e.,
\begin{equation}
{S}_\mathrm{u}^{\mathrm{mod}}(\dot{q}_{\mathrm{loss}}(r))=S_\mathrm{u}(r)
\label{eq:qloss}
\end{equation}
An example of the inferred heat loss rate, $\dot{q}_\mathrm{loss}(r)$, is shown in Fig. 4(c).

\subsection{Regime Diagram of the Observed Flame Modes}

Despite variations in the flame shape and the rotation frequency, the rotating flame phenomenon was observed over a relatively wide range of equivalence ratios and flow rates. Fig. 5 shows representative images of rotating flames at different flow rates. At relatively low mass flow rates, the rotating flames contained a single isolated wave. The flame rotation frequency increased with the mass flow rate, until a point where a second wave emerged along the edges. Multiple flame waves of approximately equal spacing appeared at higher mass flow rates, and the flame tails generally shortened as the number of waves grew with increasing flow rates. The change in the number of flame waves exhibited a certain level of hysteresis because it was affected by the instantaneous thermal boundary condition. Therefore, the boundary between single and multiple flame waves appeared fuzzy and was not recorded in the present study. However, a consistent trend was observed that for a fixed equivalence ratio, the product of the flame rotation frequency and the number of waves was seen to increase with the mass flow rate. 

\color{black}
An upper bound for the number of flame waves can be derived from the minimum flow rate requirement to sustain each isolated wave. In principle, for an N-wave rotating flame, the total flow rate ${\dot{V}}_{in}$  must exceed N times the minimum flow rate of a single-wave rotating flame ${\dot{V}}_{min}$, i.e., ${\dot{V}}_{in} \geq N{\dot{V}}_{min}$. For example, at $\phi$ = 1.25 and $H$ = 3.0 mm, the value of ${\dot{V}}_{min}$, estimated from the experimentally observed regime transition boundary, is approximately 3.4 SLPM. Therefore, a conservative lower bound for the maximum flow rate of single-wave rotating flames is approximately 6.8 SLPM. Above this value, the formation of two or more rotating flame heads is possible but depends on the specific time-histories of wall temperature and flow perturbations.
\color{black}

\begin{figure}[ht]
\includegraphics[width=0.75\linewidth]{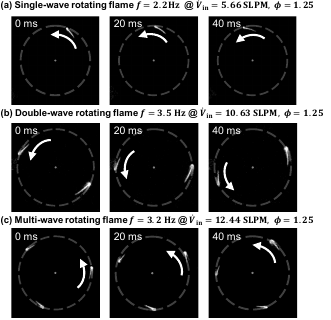}
\caption{Representative image sequences of (a) single-wave, (b) double-wave, and (c) multi-wave rotating flames at $\phi=1.25$ and a gap distance of 3.0 mm. \color{black}Multi-wave flames denote states with three or more waves.\color{black}}
\label{fig_5} 
\end{figure}

At sufficiently high flow rates, the rotating flames finally transitioned into stable ring-shaped flames anchored on the edges of the Hele-Shaw cell. A parametric study was conducted for equivalence ratios between 1.10 and 2.05, total mass flow rates between 0.94 and 22.42 SLPM, and gap distances of 2.5, 3.0 and 3.5 mm. The regime diagrams of different flame modes and their transition boundaries are presented in Fig. 6. For the aforementioned rotating flames, the widest range of mass flow rates and equivalence ratios was observed at a gap distance of 3.0 mm, and the range of mass flow rates varied monotonically with the equivalence ratio at all gap distances.

\begin{figure}[ht]
\includegraphics[width=1\linewidth]{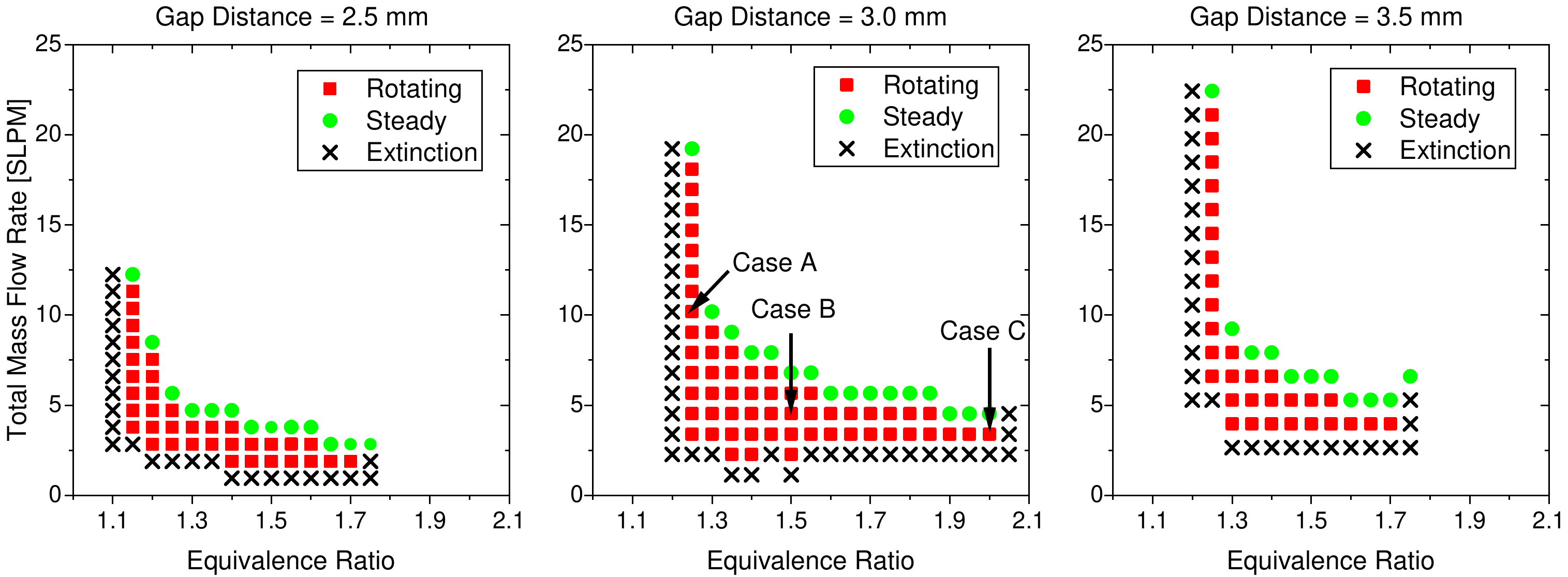}
\caption{Regime diagram of the observed flame patterns of CH$_4$-air mixtures. Numerical simulations were conducted for three selected cases in the middle diagram.}
\label{fig_6} 
\end{figure}

\begin{figure}[ht!]
\includegraphics[width=0.825\linewidth]{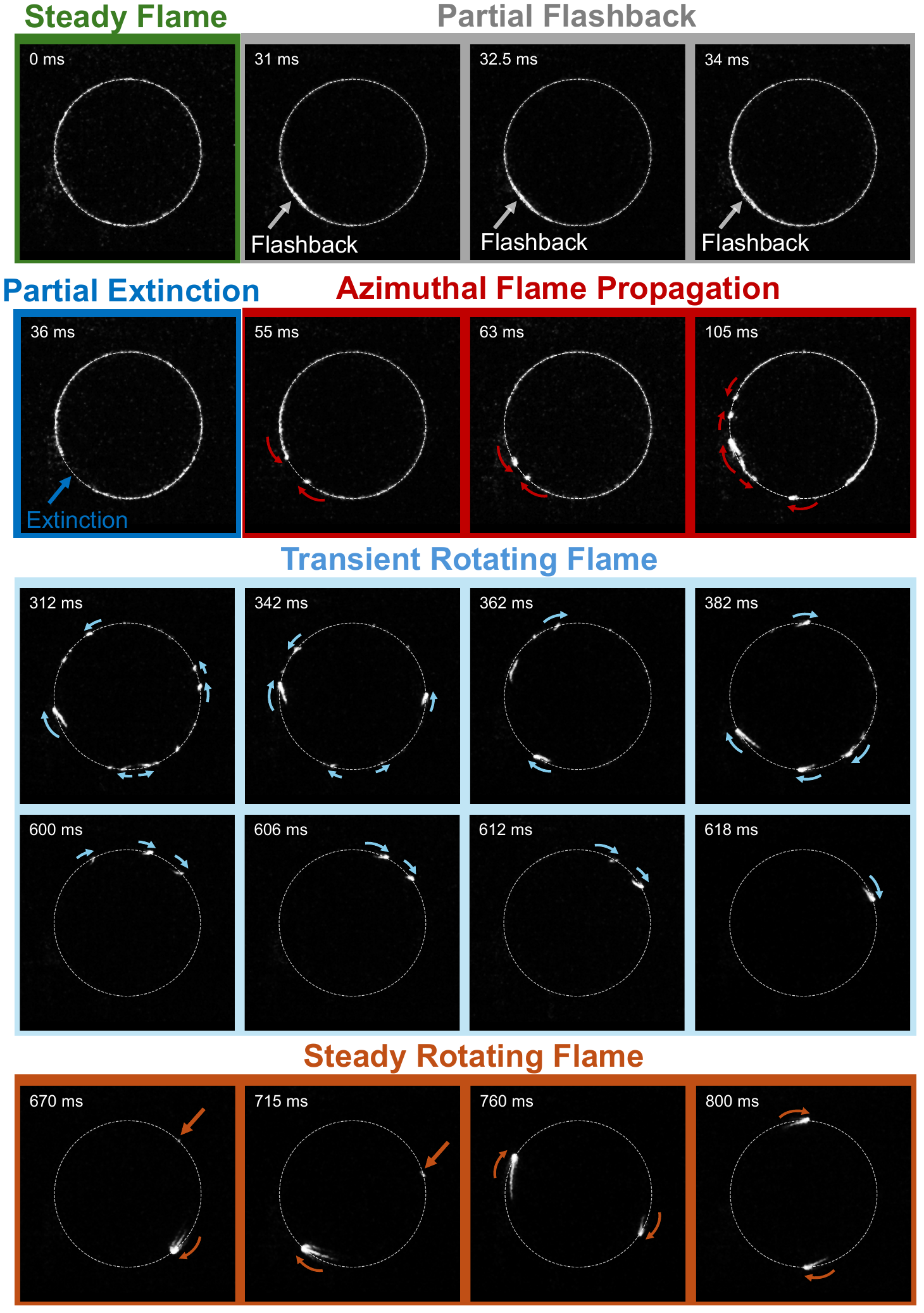}
\caption{\color{black}Representative time-resolved OH* chemiluminescence images showing the transition from a steady ring-shaped flame to a double-wave rotating flame at $\phi$ = 1.25 and $\dot{V}_{in}$= 14.70 SLPM. The sequence was recorded after switching from steady-flame conditions at a higher flow rate to the present flow rate while maintaining the same equivalence ratio.\color{black}}
\label{fig_07} 
\end{figure}

It is worth noting that the rotating flames were observed only under fuel-rich conditions. \color{black} This dependence on the equivalence ratio differs from that observed in previous studies with heated radial microchannels \cite{maruta2011micro, kumar2007formation, kumar2007pattern}, where stable rotating flames of lean methane–air mixtures could also be sustained. In those systems, the effects of wall quenching were partially counteracted by external heating. In the unheated Hele-Shaw configuration investigated in the present study, stabilization against wall quenching can instead be achieved through staged heat-release zones, such as the multi-branch structure of the rich rotating flame (see Fig. 3). \color{black} Under rich conditions, a premixed flame is confined to the interior gap of the Hele-Shaw cell, but the excess fuel in burned gas can access and react with the ambient air near the cell edges, creating a second diffusion flame that connects with the premixed flame at the rim of the cell. This diffusion branch counteracts the wall heat loss and stabilizes the entire flame structure along the radial direction of the cell. However, without any physical confinement in the tangential direction, the flame can still rotate angularly along the edges of the cell. 

\color{black}
The physical mechanism of rotating flame formation is further illustrated by the transition dynamics between steady and rotating flames. Fig. 7 shows a set of representative OH* chemiluminescence images captured during the transition from a steady ring-shaped flame to a double-wave rotating flame, as the mass flow rate was reduced from 20.00 to 14.70 SLPM while keeping a constant equivalence ratio. The steady ring-shaped flame was sustained by a balance between the local flow velocity and the effective flame speed, which were modulated by flow expansion and heat loss near the edge of the Hele-Shaw burner, respectively. At sufficiently low flow rates, heat loss at the burner edge alone could no longer sustain this balance, causing the flame to recede into the narrow gap of the burner. This recession, or flashback, exposed the flame to significantly greater heat loss and rendered its propagation unstable due to strong thermal quenching. Flashback along the burner rim readily led to local extinction.

The unburned mixture accumulated in the local extinction region until the flame front propagated azimuthally along the edge to consume it. At early times, the flame front could intermittently reattach to the burner edge, but as the wall temperature gradually decreased due to reductions in the average heat release rate, local extinction occurred more frequently, and azimuthal propagation of isolated flame waves dominated. Because the local extinction events were triggered by random fluctuations, the rotation direction and number of flame waves were inherently stochastic. However, usually after a short period of self-adjustment, the rotating flame pattern could become relatively stable and robust against minor perturbations in the flow conditions. Further details of the edge stabilization mechanism are discussed in the next section.
\color{black}

\subsection{Edge Stabilization Mechanism of the Rotating Flames}

The edges of the cell play an important role in stabilizing the rotating flames: rapid flow expansion near the edges creates a strong negative velocity gradient in the radial direction, which helps stabilize the flame against blow-off; flashback is also hindered by strong thermal quenching inside the Hele-Shaw cell. The rotating flame structure is thus sustained by a dynamic balance between the local flame speed and the flow velocity.

\begin{figure}[ht]
\includegraphics[width=\linewidth]{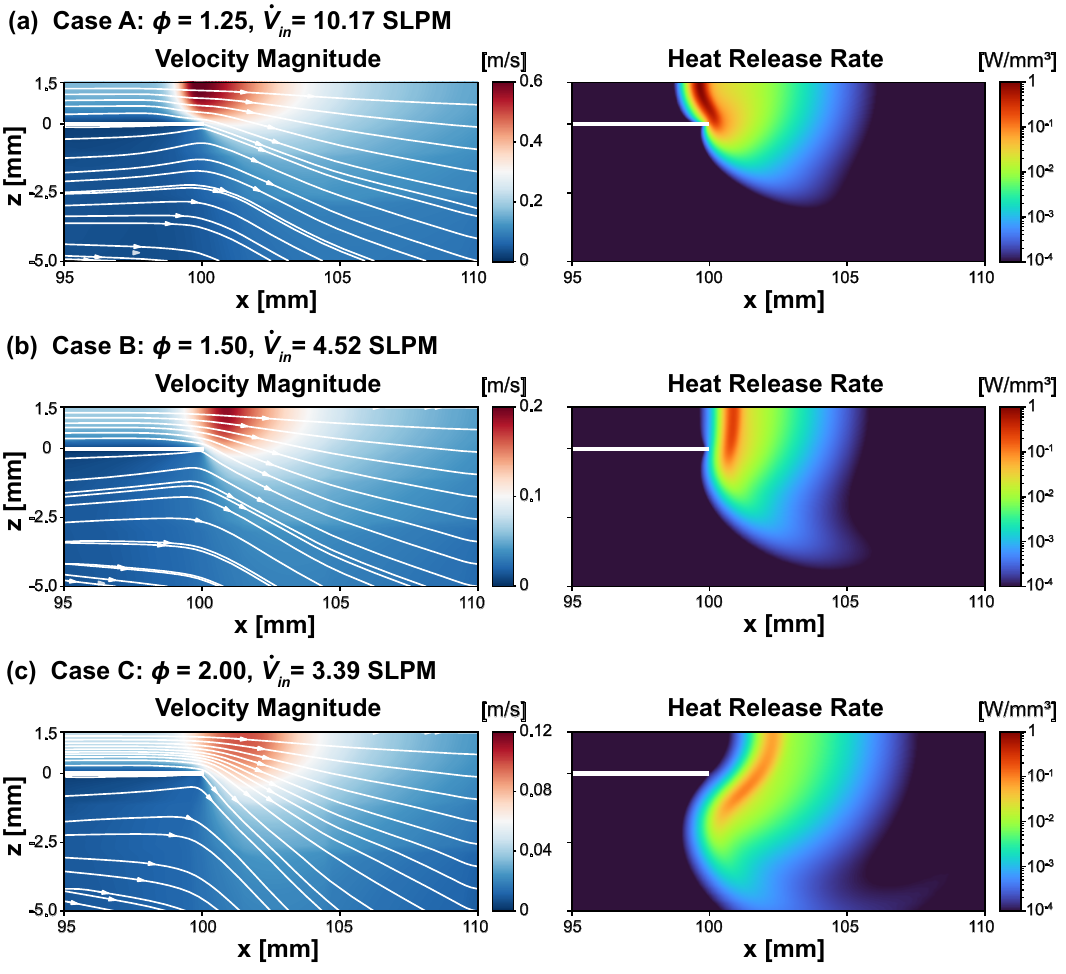}
\caption{Cross-sectional view of the edge-stabilized rotating flame structure: (a) Case A, $\phi$ = 1.25, $\dot{V}_\mathrm{in}$ =10.17 SLPM; (b) Case B, $\phi$ = 1.50, $\dot{V}_\mathrm{in}$ = 4.52 SLPM; (c) Case C, $\phi$ = 2.00, $\dot{V}_\mathrm{in}$ = 3.39 SLPM. For each case, the spatial distributions of velocity magnitude (left) and heat release rate (right) near the flame wave are presented. All results are simulated using the EBIdnsFOAM solver within the OpenFOAM platform, with detailed reactions modeled using the DRM-19 reduced mechanism.}
\label{fig_7} 
\end{figure}

The local flame structure and flow field near the edges are examined using complementary numerical simulations. \color{black} Three representative cases of 2D simulations are shown in Fig. 8. All simulations are performed across a vertical plane containing the flame tip where the premixed and diffusion branches merge. At the tip position, the azimuthal motion of the rotating flame has little effect on its stability against flashback or blow-off, as it is mainly governed by the radial velocity component. \color{black} Case A corresponds to a rotating flame under near-stoichiometric conditions. In this case, the adiabatic laminar flame speed is relatively high, and the flame tends to flash back; as illustrated in the heat release-rate distribution, the flame front is inclined inward toward the interior of the cell. Heat loss near the cell edges plays a dominant role in stabilizing the flame structure, as it reduces the local flame speed to match the flow velocity. Case B corresponds to moderately rich conditions ($\phi$ = 1.50), where the adiabatic laminar flame speed becomes comparable to the local flow velocity, and the flame front appears nearly vertical. From the heat release-rate distribution, a secondary reaction front of much weaker strength is also observed downstream, corresponding to a diffusion flame of the excess fuel burning in the ambient air. The additional heating by this diffusion flame compensates part of the heat loss and helps stabilize the flame. Case C corresponds to a rotating flame near the rich limit ($\phi$ = 2.00), where the flame tends to blow off due to its low adiabatic flame speed. In this case, the flame front is pushed outside the cell, and rapid flow expansion near the edges is the dominant stabilizing factor.

To better elucidate the edge stabilization mechanism, we have also analyzed the sensitivities of the heat-release rate ($\dot{q}$) to the total mass flow rate ($\dot{V}_{in}$) and to the equivalence ratio ($\phi$). The sensitivities are computed from the following equations:
\begin{equation}
S_V(\mathbf{x}) \;=\; \frac{\dot{q}(\mathbf{x};1.01\dot{V}_{in},\phi) - \dot{q}(\mathbf{x};0.99\dot{V}_{in},\phi)}{0.02},
\end{equation}
\begin{equation}
S_\phi(\mathbf{x}) \;=\; \frac{\dot{q}(\mathbf{x};\dot{V}_{in},1.01\phi) - \dot{q}(\mathbf{x};\dot{V}_{in},0.99\phi)}{0.02}.
\end{equation}
In these equations, $\dot{q}(\mathbf{x};\dot{V}_{in},\phi)$ represents the local heat release rate at the cross-sectional location $\mathbf{x} = (r,z)$, calculated for the specified $\dot{V}_{in}$ and $\phi$. The sensitivity ($S_V$ or $S_\phi$) is defined as the ratio of the change in $\dot{q}(\mathbf{x})$ to the relative change in the corresponding control parameters ($\dot{V}_{in}$ or $\phi$), and is numerically evaluated using a central-difference scheme that perturbs $\dot{V}_{in}$ or $\phi$ by $\pm 1\%$. The results are displayed in Fig. 9.

\begin{figure}[ht]
\includegraphics[width=0.9\linewidth]{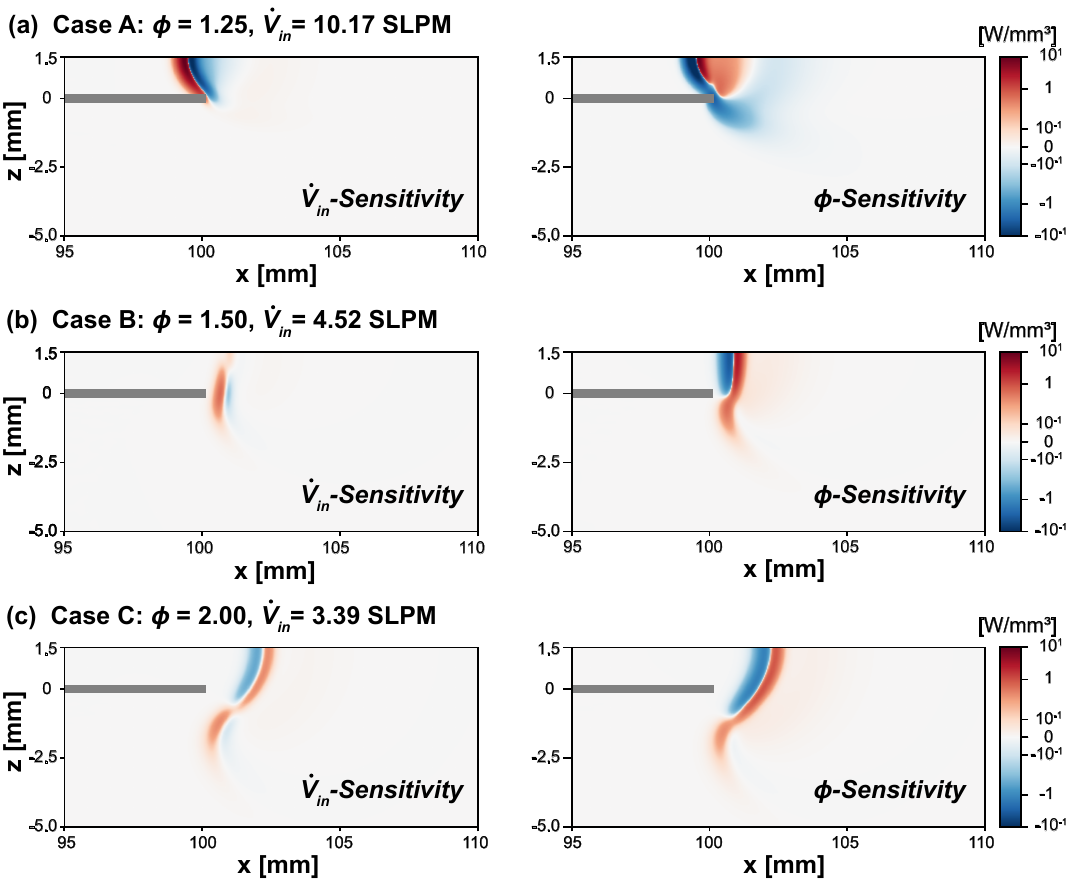}
\caption{Sensitivities of the local heat-release rate to the total mass flow rate (left) and to the equivalence ratio (right). Results are presented for all three cases shown in Fig. 8.}
\label{fig_9} 
\end{figure}

In Case A, the sensitivity analysis reveals a counterintuitive phenomenon: increasing the flow rate causes the flame front to recede toward the interior of the cell rather than advance toward the exterior. This likely results from the increase in thermal power counteracting heat loss and accelerating the local flame speed. A similar trend is observed in Case B, although the magnitude of the sensitivity is weaker. A closer examination also reveals that the sign of sensitivity reverses near the center of the gap (1.0 mm $\le z \leq$ 1.5 mm), suggesting that the center region of the flame wave is stabilized by a different mechanism (flow expansion instead of wall heat loss). In Case C, the majority of the flame front (0 mm $\le z \leq$ 1.5 mm) protrudes outward as the flow rate increases and is stabilized primarily by the flow expansion; in contrast, the bottom portion of the flame front ($z \le$ 0 mm) tilts inward and is primarily stabilized by wall heat loss. In all three cases, increasing the equivalence ratio reduces the laminar flame speed and pushes the flame front position downstream. Note that Cases A and C are very close to their stability limits: a further decrease in $\phi$ in Case A would likely trigger flashback, while a further increase in $\phi$ in Case C would likely cause blow-off.

\color{black}
An additional full 3-D simulation over a 360-degree computational domain was conducted for the representative single-wave case at $\phi=1.25$ and $\dot{V}_{\rm in}=10.17$ SLPM. The resulting instantaneous distribution of the heat release rate is shown in Fig. 10. A flame head is observed near the burner edge, followed by an elongated trailing flame that extends into the gap. The simulated rotation frequency is approximately 2.5 Hz, which is close to the measured value of 3.0 Hz under the same conditions.
\color{black}

\begin{figure}[ht]
\includegraphics[width=0.55\linewidth]{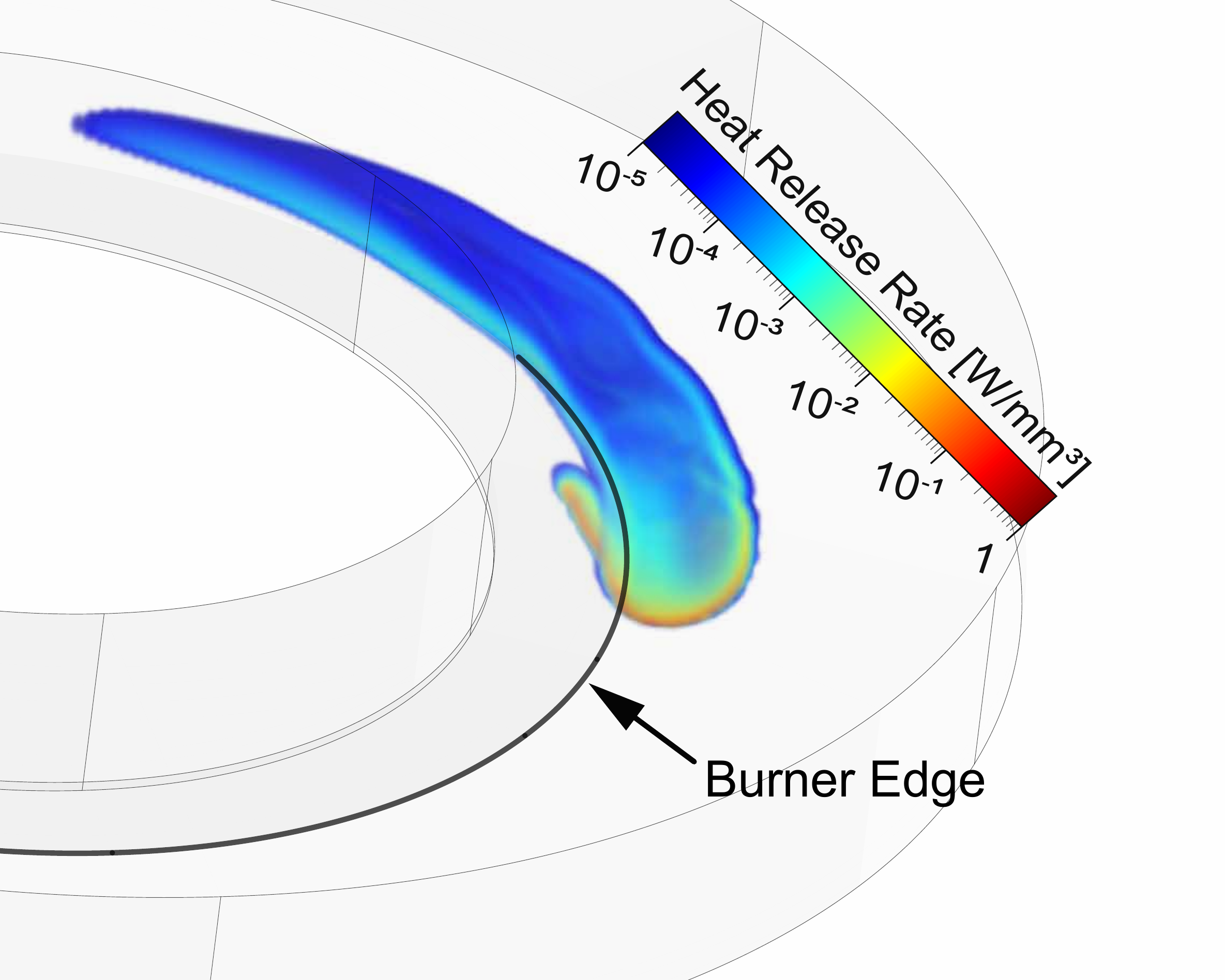}
\caption{Three-dimensional simulation results for the single-wave rotating flame at $\phi$ = 1.25 and $\dot{V}_{in}$ = 10.17 SLPM over a 360-degree computational domain.}
\label{fig_10} 
\end{figure}

\subsection{\color{black}Semi-Empirical Model for the Single-wave Rotating Flames\color{black}}

For the single-wave rotating flames, we have formulated a semi-empirical model that retains the essential geometric and kinematic features of the rotating front. Note that the aim of this model is not to provide \textit{ab initio} predictions of detailed flame structure and rotation frequency, but to establish quantitative relations with global control parameters such as $\dot{V}_{in}$, $\phi$ and $T_s$, based on simple assumptions.

\begin{figure}[ht]
\includegraphics[width=0.5\linewidth]{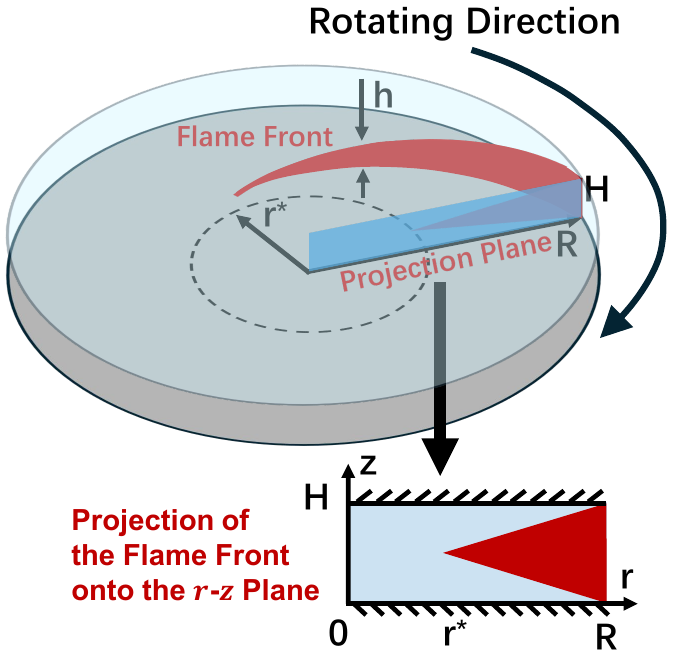}
\caption{Schematic of the \color{black}semi-empirical \color{black}representation of a rotating flame surface}
\label{fig_11} 
\end{figure}

In the present model, a rotating flame is represented as a curved surface that occupies an annular region of the cell with radial extent $r^* \le r \le R$ (where $r^*$ denotes the minimum radial location of the flame tail), as illustrated in Fig. 11. At steady state, a balance is established between the supply and consumption rates of fresh mixture. In the co-rotating frame of reference, this balance is expressed by the following equation:
\begin{equation}
\dot{V}_{\mathrm{in}} = \iint_{\mathrm{flame}} \mathbf{n} \cdot \mathbf{v} \, \frac{T_0}{T_s} \, dS
= \frac{2}{3}\int_{r^*}^{R} h(r) S_\mathrm{u}(r) \frac{T_0}{T_s(r)} \sqrt{1+(r/r')^2}\,dr,
\label{eq:Vin}
\end{equation}

The right-hand side of this equation is evaluated along the central plane of the cell, where the factor 2/3 represents the ratio between the mean and maximum velocity in Poiseuille flow. $h$ denotes the effective flame height in the gap of the Hele-Shaw cell, which is approximated as a linear function of the radial location $r$:
\begin{equation}
h(r)=H \cdot \frac{r-r^*}{R-r^*},
\label{eq:h}
\end{equation}

Substituting Eqs. (6), (7), (8) and (13) into Eq.(12) yields:
\begin{equation}
\begin{split}
\dot{V}_{\mathrm{in}} & = \frac{2}{3} \frac{H}{R-r^*} \int_{r^*}^{R} \bigl(r-r^*\bigr) \frac{T_0}{T_s(r)}\bigg[ v_r(r) \frac{r}{r'} - v_\theta(r)\bigg] dr \\
& = \frac{2}{3} \frac{H}{R-r^*} \int_{r^*}^{R} \bigl(r-r^*\bigr) \frac{T_0}{T_s(r)}\bigg[ \frac{3}{2} \frac{\dot{V}_{in}}{2 \pi r H} \frac{T_s(r)}{T_0}\frac{r}{r'}- 2 \pi r f\bigg] dr\\
& = \frac{\dot{V}_{in}}{2 \pi} \frac{1}{R-r^*} \int_{r^*}^{R} \frac{r-r^*}{r'} dr - \frac{4 \pi}{3} \frac{f H}{R-r^*} \int_{r^*}^{R} \frac{T_0}{T_s(r)} (r-r^*) r dr 
\end{split}
\end{equation}

As found in the present study, the spatial variations in the surface temperature $T_s(r)$ are usually small, especially in the rotating flame region (see Fig. 4). Therefore, a reasonable approximation of $T_s(r) \approx T_R$ is applied, where $T_R$ represents the surface temperature at the cell edges ($r = R$). In addition, the radial and angular spans of the rotating flame, defined as $\Delta r = R - r^*$ and $\Delta\theta = \theta(R) - \theta(r^*)$, respectively, are also small. A first-order approximation to Eq. (14) is as follows:

\begin{equation} \label{eq:qin1}
\dot{V}_{\mathrm{in}} \approx  \frac{\Delta\theta}{2 \pi} \dot{V}_{in} - \frac{2 \pi}{3} \frac{T_0}{T_R}  f H R \Delta r 
\end{equation}

The rotation frequency $f$ can be determined as:
\begin{equation}
 f = \Big( 1 - \frac{\Delta \theta}{2 \pi} \Big) \frac{T_R}{T_0} \frac{3 \dot{V}_{in}}{2 \pi H R \Delta r} \approx \frac{T_R}{T_0} \frac{3 \dot{V}_{in}}{2 \pi H R \Delta r}
\end{equation}

Additional information on $\Delta r$ is needed to close this model. In the present study, this is provided by a second constraint on the flame shape $r(\theta)$, including the radial location of the flame tail $r^*$, based on heat-loss analysis. As shown in Fig. 4(c), the effective heat loss rate $\dot{q}_{\mathrm{loss}}$ is roughly linear with respect to $r$. Hence, a simple approximation is adopted:
\begin{equation}
\dot{q}_\mathrm{loss} = \kappa (R-r),
\label{eq:r_loss}
\end{equation}
where $\kappa$ is an effective sensitivity parameter that depends on the $\dot{V}_\mathrm{in}$ and $T_R$. In the present study, an empirical expression for $\kappa$ is fitted from a subset of the experimental data (about 20\% of the total data): $\kappa(\dot{V}_\mathrm{in},T_R) = a_0+a_{1}\dot{V}_\mathrm{in}+a_{2}(T_R-T_0)$, where $a_0=2410$ W/mm$^4$, $a_1=-79.9$ W/mm$^4$-SLPM, $a_2=11.8$ W/mm$^4$-K and $T_0$ = 300 K.

Fig. 4 also shows that quenching of the flame tail occurs at a local flame speed of approximately 40\% of the adiabatic value. In accordance with this observation, we introduce a critical heat loss rate ($\dot{q}^*_\mathrm{loss}$) that satisfies:

\begin{equation}
S_\mathrm{u}^{\mathrm{mod}}(\dot{q}^*_\mathrm{loss}; T_R)=0.4 \, S_\mathrm{u}^{\mathrm{ad}}(T_R),
\label{eq:eta}
\end{equation}
where $S_\mathrm{u}^{ad}(T_R)$ is the adiabatic laminar flame speed evaluated at $T_R$. Once $\dot{q}^*_\mathrm{loss}$ is determined from Cantera simulations, the flame tail location $r^*$ can be calculated as follows:

\begin{equation}
r^* = R - \dot{q}^*_\mathrm{loss}/\kappa
\label{eq:rtip}
\end{equation}

Subsequently, the flame rotation frequency $f$ can be obtained by substituting the value of $\Delta r = R-r^*$ into Eq. (16). In addition, the flame shape $r(\theta)$ can be determined by numerically solving an ordinary differential equation in polar coordinates:

\begin{equation}
S_\mathrm{u}^{\mathrm{mod}}(\dot{q}_\mathrm{loss}(r)) = \frac{v_r(r)\,r - v_\theta(r) \,r'}{\sqrt{r^2+(r')^2}} = \frac{\frac{3 T_R}{2 T_0} \frac{\dot{V}_{in}}{2 \pi H} - 2 \pi r r'f}{\sqrt{r^2+(r')^2}}.
\label{eq:flame shape}
\end{equation}

\begin{figure}[ht]
\includegraphics[width=\linewidth]{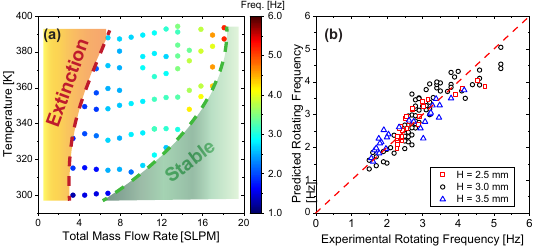}
\caption{Rotation frequency of the observed single-wave rotating flames at $\phi$=1.25. (a) Measured rotation frequency as a function of  $\dot{V}_\mathrm{in}$ and $T_R$ \color{black}for gap distance $H$ = 3.0 mm. \color{black}(b) Comparison between the predicted flame rotation frequencies from the semi-empirical model with the measured values \color{black}for $H$ = 2.5, 3.0 and 3.5 mm. The empirical parameter $\kappa$ was calibrated using only a subset of the $H$=3.0 mm data\color{black}}
\label{fig_12} 
\end{figure}

The performance of this model is validated against direct experimental measurements of the flame rotation frequencies over 3.4 SLPM $\le \dot{V}_{in} \le$ 18.1 SLPM and 300 K $\le T_R \le$ 394 K and at a fixed equivalence ratio of $\phi$ = 1.25. The results are shown in Fig. 12. The measured rotation frequencies range from 1.3 to 5.2 Hz and show a positive dependence on both $\dot{V}_{in}$ and $T_R$. Flame extinction occurs at sufficiently low $\dot{V}_{in}$, with a transition value between 3.4 and 6.8 SLPM that increases with $T_R$. Stable ring-shaped flames are observed at sufficiently high $\dot{V}_{in}$, and the transition value also increases with $T_R$. \color{black} To examine the generalizability of the model with respect to gap distance, additional single-wave rotating-flame data at gap distances $H$ = 2.5 mm and $H$ = 3.5 mm are included in Fig. 12(b). The empirical parameter $\kappa$ is calibrated using only a subset (approximately 20\%) of the $H$ = 3.0 mm data, and no additional fitting is performed for other cases. As shown in Fig. 12(b), the predicted flame rotation frequencies agree reasonably with the measured values for all three gap distances. Extension of the present model to broader ranges of gap distances and equivalence ratios is planned for future studies.\color{black}

\section{Conclusions}
Self-sustained propagation of edge-stabilized rotating flames was observed in an unheated, open circular Hele-Shaw cell for rich CH$_4$-air mixtures. As revealed by OH-PLIF measurements, these flames exhibit a bibrachial flame structure, with a diffusion branch gliding along the side edges and a premixed branch extending into the interior of the cell. Complementary simulations indicate that rotating flames arise from a dynamic balance between the local flame speed and the unburned-gas velocity near the edges, with both heat loss and flow expansion playing key roles in stabilizing the rotation pattern. Heat loss through the walls reduces the local flame speed and provides resistance against flashback, whereas rapid flow expansion near the cell edges creates a zone of negative velocity gradient that delays blow-off. A parametric study is conducted for various equivalence ratios, flow rates, and gap distances, from which the regime diagrams of flame modes and rotation frequencies are obtained. At low flow rates, the rotating state is single-wave, while increasing flow rate promotes multi-wave rotating patterns and eventually a transition to stable ring-shaped flames on the cell edges. \color{black}For single-wave rotating flames at $\phi=1.25$, a semi-empirical model is established to predict their frequencies and shapes as functions of the total flow rate and the surface temperature at the burner edge. \color{black}Results of the current study should be useful to the advancement of micro-combustion technologies as well as fundamental research of laminar flame dynamics.

\begin{acknowledgments}
This work was supported by the National Natural Science Foundation of China under Grants No. 12472278, the National Key Research and Development Program of China under Grant No. 2025YFF0511801, \color{black} and the Open Research Program of National Key Laboratory of Fundamental Algorithms and Models for Engineering Simulation. \color{black} Numerical simulations were also supported by the High-Performance Computing Platform of Peking University.
\end{acknowledgments}

\section*{Data availability statement}
The data that support the findings of this article are openly available \cite{Nie2026}.

\bibliography{References}

\end{document}